\documentclass{PoS}

\newcommand{\gev}{\ensuremath{\mathrm{\;GeV}}}

\title{Diffractive Structure Functions with H1}

\ShortTitle{Diffractive Structure Functions with H1}

\author{\speaker{Paul Laycock}\\
        University of Liverpool\\
        E-mail: \email{laycock@hep.ph.liv.ac.uk}}

      \abstract {H1 has measured the diffractive DIS cross section $ep
        \rightarrow eXY$ using data from both of the HERA data-taking
        periods.  Using new measurements of the diffractive cross
        section at different centre-of-mass energies, the diffractive
        longitudinal structure function $F_L^D$ has been extracted.
        The results are in agreement with NLO QCD predictions based on
        fits to inclusive data.  New high statistics measurements of
        the diffractive reduced cross section $\sigma_r^D$ have been
        made using two experimental methods covering the accessible
        kinematic range.  This precise dataset agrees well with
        QCD-based predictions.}

\FullConference{The 2011 Europhysics Conference on High Energy Physics-HEP 2011,\\
                July 21-27, 2011\\
                Grenoble, Rh\^one-Alpes France}

\begin{document}

\section{Introduction}
The diffractive DIS process $ep \rightarrow eXY$ factorises in QCD.
An additional assumption is often made whereby the proton vertex
dynamics factorise from the vertex of the hard scatter - proton vertex
factorisation.  Although proton vertex factorisation must be broken in
QCD, measurements show that it's a good enough approximation to the
data such that meaningful next-to-leading order (NLO) QCD fits can be
made~\cite{H1:LRG}.

In analogy with the inclusive case, the DDIS cross section can be
expressed in terms of a linear combination of structure functions.  In
the HERA kinematic regime, this can be well approximated by $F_2^D$
and a term related to scattering of longitudinally polarised photons
$F_L^D$.  

Diffractive events are selected either by detecting the final state
proton, or on the basis of a Large Rapidity Gap (LRG) being present.
In the latter case, the final state system $Y$ escapes detection, the
cross section is integrated over ranges in leading baryon mass $M_Y$
and $t$.

\section{Structure Function Measurements}
Data from three proton beam energies, $E_p=460, 575$ and $920$ GeV,
have been used to measure the diffractive reduced cross section at the
same $x$ and $Q^2$, but different $y$.  Following a similar procedure
to that used for the extraction of $F_L$ by H1, these data have then
been used, together with previously published data at $820$
GeV~\cite{H1:LRG}, to extract the diffractive longitudinal structure
function $F_L^D$.

\begin{figure}
  \begin{center}
    \includegraphics[width=0.65\columnwidth]{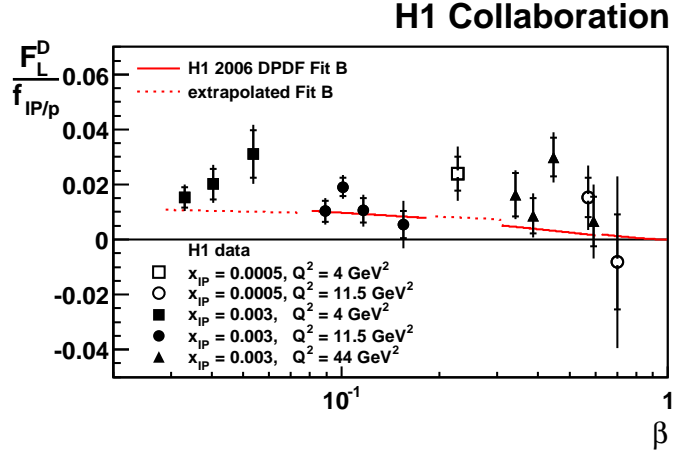}
    \caption{The measurements of $F_L^D$, as a function of $\beta$.
      The data are compared to the predictions of Fit B, the
      extrapolation of this fit is shown as a dashed
      line.}\label{Fig:FLD}
  \end{center}
\end{figure}

In order to extract $F_L^D$ optimally, the cross sections are
normalised to the H1 2006 DPDF Fit B result.  As the published data at
$820 \gev$ were included in the analysis of the data used as input to
H1 2006 DPDF Fit B, they are already consistently normalised.  The
resulting measurements of $F_L^D$~\cite{FLD} are shown in figure
$\ref{Fig:FLD}$.  The data agree well with the predictions of Fit B.

\begin{figure}
  \begin{center}
    \includegraphics[width=0.9\columnwidth]{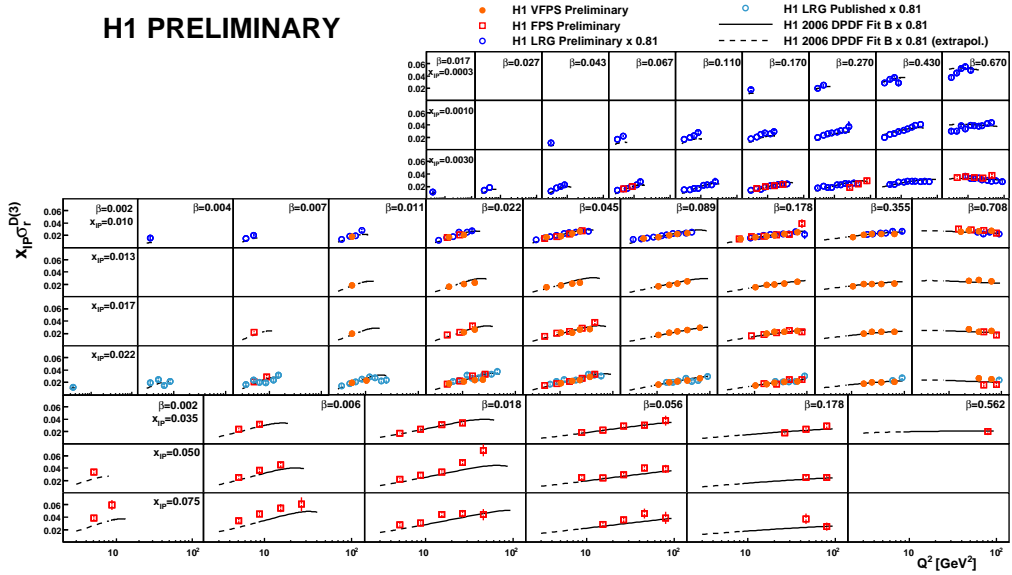}
    \caption{The measurements of $\sigma_r^D$ using all experimental
      methods.  }\label{Fig:F2D}
  \end{center}
\end{figure}

Data at the nominal proton beam energy of $E_p=920$ GeV have been
analysed to extract the diffractive reduced cross section $\sigma_r^D$
in as wide a kinematic range as possible, using both the tagged proton
and the LRG methods.  These new, precise data~\cite{F2D:Prelim, FPS,
  VFPS:Prelim} agree well with the published H1 data~\cite{H1:LRG} and
cover almost the entire accessible kinematic range.  The full set of
measurements are shown in figure $\ref{Fig:F2D}$.  The data are shown
as a function of $Q^2$ in bins of $\beta$, both the LRG data and Fit B
have been normalised to $M_Y=M_{proton}$.  The data compare well with
Fit B.

\section{Conclusions}
New measurements of the diffractive reduced cross section using data
taken at three proton beam energies have been made.  The measurements
have been combined with existing H1 data at $820$ GeV in order to
extract $F_L^D$.  The result agrees well with the predictions of Fit
B.  New measurements of $\sigma_r^D$ using both the proton tagged and
LRG methods have also been made, which cover nearly the whole
accessible kinematic range.  The data agree well with published data
and with Fit B.


\end{document}